\documentclass[10pt,journal,compsoc]{IEEEtran}

\ifCLASSOPTIONcompsoc
  \usepackage[nocompress]{cite}
\else
  \usepackage{cite}
\fi
\usepackage{amsmath}
\usepackage[ruled,linesnumbered]{algorithm2e}
\usepackage{graphicx}
\usepackage{amsfonts}
\usepackage{booktabs}
\usepackage{float}

\usepackage{array}
\usepackage{multirow}
\usepackage{longtable}
\usepackage{rotating}

\begin{document}
%
\title{Com-DDPG: A Multiagent Reinforcement Learning-based Offloading Strategy for Mobile Edge Computing}
%
%
%
%

\author{Honghao~Gao,~\IEEEmembership{Senior Member,~IEEE,}
        Xuejie~Wang,
        Xiaojin~Ma,
		Wei~Wei,~\IEEEmembership{Senior Member,~IEEE,}
		and Shahid~Mumtaz,~\IEEEmembership{Senior Member,~IEEE,}
\thanks{The research is supported by the National Key R\&D Program of China (No. 2020YFB1006003)}
\IEEEcompsocitemizethanks{
\IEEEcompsocthanksitem Honghao~Gao is with the School of Computer Engineering and Science, Shanghai University, Shanghai 200444, China and Gachon University, Gyeonggi-Do 461-701, South Korea
(e-mail: gaohonghao@shu.edu.cn.).
\IEEEcompsocthanksitem Xuejie~Wang is with the School of Computer Engineering and Science, Shanghai University, Shanghai 200444, China
(Corresponding authors E-mail: wangxuejie@shu.edu.cn.).
\IEEEcompsocthanksitem Xiaojin~Ma is with the School of Computer Engineering and Science, Shanghai University, Shanghai 200444, China \& School of Management, Henan University of Science and Technology, Luoyang 471000 China (e-mail: xjma@shu.edu.cn.)
\IEEEcompsocthanksitem Wei~Wei is with the School of Computer Science and Engineering, Xi’an University of Technology, Xi’an 710048, China 
(e-mail: weiwei@xaut.edu.cn.).
\IEEEcompsocthanksitem Shahid~Mumtaz is with the Instituto de Telecomunicaces, Aveiro, Portugal
(e-mail: Dr.shahid.mumtaz@ieee.org.).}
}


\IEEEtitleabstractindextext{
\begin{abstract}
The development of mobile services has impacted a variety of computation-intensive and time-sensitive applications, such as recommendation systems and daily payment methods. However, computing task competition involving limited resources increases the task processing latency and energy consumption of mobile devices, as well as time constraints. Mobile edge computing (MEC) has been widely used to address these problems. However, there are limitations to existing methods used during computation offloading. On the one hand, they focus on independent tasks rather than dependent tasks. The challenges of task dependency in the real world, especially task segmentation and integration, remain to be addressed. On the other hand, the multiuser scenarios related to resource allocation and the mutex access problem must be considered. In this paper, we propose a novel offloading approach, Com-DDPG, for MEC using multiagent reinforcement learning to enhance the offloading performance. First, we discuss the task dependency model, task priority model, energy consumption model, and average latency from the perspective of server clusters and multidependence on mobile tasks. Our method based on these models  is introduced to formalize communication behavior among multiple agents; then, reinforcement learning is executed as an offloading strategy to obtain the results. Because of the incomplete state information, long short-term memory (LSTM) is employed as a decision-making tool to assess the internal state. Moreover, to optimize and support effective action, we consider using a bidirectional recurrent neural network (BRNN) to learn and enhance features obtained from agents' communication. Finally, we simulate experiments on the Alibaba cluster dataset. The results show that our method is better than other baselines in terms of energy consumption, load status and latency.
\end{abstract}

\begin{IEEEkeywords}
Offloading Strategy, Multiagent Reinforcement Learning, Mobile Edge Computing, Bidirectional Recurrent Neural Network, Agent Communication Behavior
\end{IEEEkeywords}}

\maketitle

\IEEEdisplaynontitleabstractindextext

%
\IEEEpeerreviewmaketitle

\ifCLASSOPTIONcompsoc
\IEEEraisesectionheading{\section{Introduction}\label{sec:introduction}}
\else
\section{Introduction}
\label{sec:introduction}
\fi

\IEEEPARstart{M}{obile} devices, which allow computing and communication at anytime and any where, are considered key to pervasive computing, promoting the prosperous mobile industry~\cite{shi2016edge,satyanarayanan2010mobile}. However, computation-intensive and time-sensitive applications of the mobile Internet call for flexible computing means to address the explosive growth of data generated and used by mobile devices for tasks such as image processing, video streaming, and AR/VR data~\cite{gao2020mining,yang2019approach}. For example, artificial intelligence-based applications have high demands for computing resources. Although the cloud architecture has the ability to process big data, it faces challenges related to the impact of network speed and transmission time on user experience~\cite{yin2020qos}. By contrast, performing all these computations via mobile device may be impossible to due to limitations of resources, storage and energy consumption. If a devices, servers and the cloud work in the cooperative manner, the performance of computing tasks can be improved substantially.

In  general, a third party requests computing resources and submits tasks and data to a cloud center via the Internet~\cite{arora2013secure,jiang2019toward}. For example, mobile cloud computing was introduced to improve the capabilities of mobile devices~\cite{dinh2013survey}. This centralized method has unlimited computing resources due to tens of thousands or more servers being clustered together at the end of the cloud~\cite{Mell2010The}.  However, the cloud is usually far from mobile device, and the latency and energy consumption in transmitting big data is enormous due to the network traffic and speed. Thus, mobile cloud computing is limited by the operating efficiency. MEC and task offloading were proposed to address these issues. A MEC node can be a local server, nearby server or mobile device that provides the 'computing' functions. To ensure high-quality network services and low latency, task offloading deploys the computing task to the edge of the mobile network and provides communication, storage, and computing resources at a  local device~\cite{zhang2016energy,mao2017survey}. 

As an offloading strategy of task computing, MEC determines the target server/servers where a task or group of subtasks can be executed~\cite{flores2015mobile,jiao2013cloud,mach2017mobile}. If the nearby mobile device is idle and has the ability to provide services, it can be selected as a server, called an edge server. Furthermore, the evaluation criteria, including CPU, throughput, storage, and network bandwidth, at each MEC server are important to guide the task offloading process~\cite{mao2017stochastic}. Two main types of task offloading exist: the coarse-grained method and the fine-grained method~\cite{dinh2017offloading,kuang2020offloading}. The former considers mobile applications as a single object requiring offloading rather than dividing them into multiple subtasks. However, this method does not effectively utilize the distributed computing characteristics under the MEC environment. The latter divides mobile applications into multiple subtasks. Then, the parties or all subtasks are offloaded to multiple MEC severs for data processing and transmission. This method reduces the task time and has a high rate of resource utilization. Although the divided subtasks entail lower computational complexity and less data transmission, a data dependency problem exists among subtasks. Thus, one important question is how to handle these dependencies when configuring a strategy for offloading. Another factor is cooperation when subtasks, except the resource competition, are deployed to run at the edge server.

This paper proposes an approach, Com-DDPG, for multiagent reinforcement learning-based offloading for MEC. We aim to use reinforcement learning to encode different factors as input for the mobile environment and then output the result as the offloading strategy according to the feature learning. The innovation is the attention given to the communication behavior during multiagent reinforcement learning to address the data dependency problem. LSTM and a bidirectional recurrent neural network (BRNN) are used to improve the reinforcement learning, and LSTM is employed to predict and confirm the learning of state information because reinforcement learning is a black box process and the internal state is hard to observe. Moreover, BRNN is added to the reinforcement learning framework as a new layer to discover more communication features. The main contributions of this paper are as follows:

$ \bullet $ We introduce a multidevice and multiserver computation offloading framework for heterogeneous MEC, which is used to simulate the resource competition between multiple devices and the dependency between subtasks. The priority between tasks is also considered.

$ \bullet $ We discuss how, where and what tasks can be offloaded when using reinforcement learning for MEC. Com-DDPG minimizes the energy consumption, load status, execution latency and network usage and considers the dependency between subtasks.

$ \bullet $ We use bidirectional LSTM and BRNN as additional hidden layers to improve the offloading strategy when reinforcement learning is applied.

The remainder of this paper is organized as follows. Section 2 introduces related work. The system architecture is described in Section 3. The proposed method for task offloading is illustrated in Section 4. Section 5 discuss the experiments and results. Finally, Section 6 concludes the paper and discusses directions for future research.

\section{Related work}
The key factor to achieve high performance in computation offloading scenarios is an effective offloading algorithm. According to different optimization objectives and scheduling strategies, this section briefly compares research on task offloading in MEC. Various studies have investigated the MEC offloading from different perspectives and measured the performance of the algorithm based on energy consumption, computing capability, and resource utilization rate. Some studies have attempted to extend the single-user offloading problem to a multiuser offloading problem to adapt to real-world challenges.

Aiming to minimize the energy consumption of mobile terminal devices or to make a tradeoff between energy consumption and delay according to the needs of different tasks. 
Mao Y et al.~\cite{mao2016dynamic} proposed the LODCO algorithm, a dynamic computing offloading algorithm based on Lyapunov optimization theory. This method optimizes the offloading decision from execution latency and task failure. The algorithm is used to minimize the offloading task processing delay and guarantees success in the data transmission process, thereby reducing the chance of offloading failure. 
Cui Y et al.~\cite{cui2019resource} proposed an intelligent offloading and resource allocation algorithm for a multitype offloading platform. The K-means algorithm is used to select a platform to offload, and reinforcement learning is applied to solve computational resource allocation problems in nonlocal computing.
Ali Z et al.~\cite{ali2019deep} proposed an efficient energy-saving computational offloading scheme based on deep learning  to solve the selective offloading of mobile application components. The cost function was formulated in terms of residual energy, computing load, energy consumption, amount of transmitted data and communication delay to determine the cost of all possible combinations of component offloading strategies. Furthermore, a deep learning network was trained to provide alternatives for the extensive computations. Experimental results showed that the scheme has high accuracy and low energy consumption. 
Ning Z et al.~\cite{ning2018cooperative} combined MCC and MEC to make computation offloading decisions. Considering the rich computing resources of MCC and the low transmission delay of MEC, the iterative heuristic MEC resource allocation (IHRA) algorithm was proposed to offload computing tasks to the MEC server or MCC server in multiuser situations. The author expanded the single-user offloading problem to the multiuser offloading problem, while considering resource constraints and interference among multiple users.
Nan Y et al.~\cite{nan2017adaptive} presented a solution combining Lyapunov optimization theory with an adaptive online learning method for optimal offloading to consider the trade-off between response delay and energy consumption in the context of the Internet of things. 

Consider the potential service congestion caused by multiuser competition for computing resources. 
Guo B et al.~\cite{gu2019task} considered a multiuser MEC system in which one MEC server handles the computing tasks that multiple users offloads via wireless channels. The total delay and energy consumption are used as offloading indicators. Furthermore, a multiuser MEC system was proposed to minimize the total cost of the considered MEC system. Based on the multiuser MEC system model, the author established a network model, task model and computing model and modeled the offloading problem in the multiuser MEC system as an optimization problem. To solve this problem, the author proposed solutions based on Q-Learning and deep Q-learning.
Li J et al.~\cite{li2018deep} studied the multiuser service delay problem in MEC offloading scenarios and proposed a partial computation offloading model. An optimization strategy was used to optimize the communication and computing resource allocation. The experiment was conducted in a specific scenario where the communication resources are much larger than the computing resources. Compared with the local execution and edge execution of  tasks, the proposed partial offloading strategy minimizes the total delay. 
J. Ren et al.~\cite{ren2017partial} considered the multiuser service latency in MEC and presented a partial computational offloading policy to optimize communication and computing resource allocation. Experiments were performed in specific environment with sufficient network bandwidth. The proposed scheme reduced device latency  and improved the quality of service (QoS) for users.
Cao H et al.~\cite{cao2017distributed} described the multiuser computation offloading decision problem as a noncooperative game. To maximize the utility function, consisting of the communication cost and the calculation cost of offloading, the author presented a fully distributed computation offloading scheme (FDCO) based on machine learning technology. 
Teng Ying-lei et al.~\cite{han2020joint} optimized the multiuser mobile edge computing and offloading system, constructed a Markov decision problem with  time delay and long-term average power consumption objectives, and solved the problem via convex optimization theory. 

Heuristic algorithms and meta-heuristic algorithms are widely used to solve NP-hard problems, such as task offloading, but both approaches have shortcomings. Heuristic algorithms easily fall into local minima, and the overall optimal result is difficult to obtain. Meta-heuristic algorithms have an excessive number of parameters, the calculation results are difficult to reuse, and parameter tuning cannot be performed quickly and effectively. In contrast, deep reinforcement learning combines the advantages of deep learning and reinforcement learning and has the characteristics of self-learning and self-adaptation~\cite{jin2020provably}. However, the results of deep reinforcement learning algorithms rely on complete state information and ignore the cooperation among multiple users. Therefore, the DDPG algorithm implements an LSTM network and multiagent collaboration to overcome the defects of deep reinforcement learning, and the agent makes decisions independently during the training process, thereby solving the problem of MEC offloading tasks with a large number of mobile devices.

\begin{figure*}
	\centering
	\includegraphics[scale=0.6]{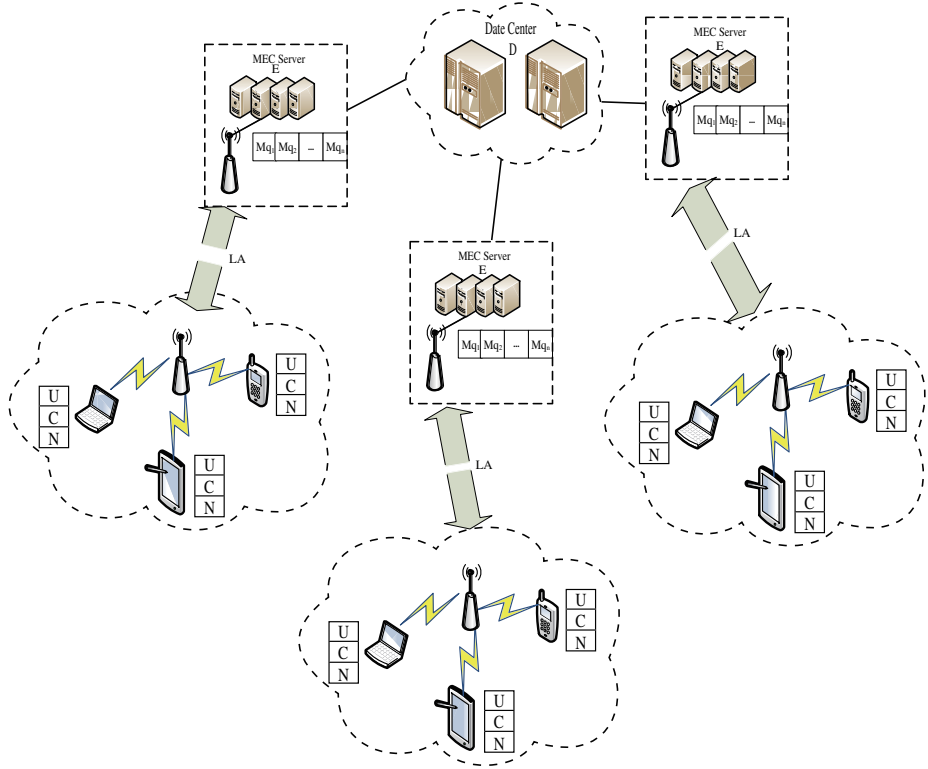}
	\caption{The framework of MEC.} \label{struct}
\end{figure*} 

\section{System Model}
In this section, the system model of our offloading framework is introduced. Then, the basic notation used in the study is presented.  

As shown in Fig.~\ref{struct}, the framework consists of mobile devices, edge servers and cloud data centers. The mobile device level $ M $ includes devices with low processing performance,  such as tablets, laptops and mobile phones. The cloud data center level $ D $  is a cluster that contains a large number of high-performance servers. The edge level $ E $ is different from the cloud center. The main feature is being located near the user side or data generation side. The server belonging to the edge level is divided into several regions according to performance and relative distance. For the MEC environment, mobile devices transmit computing tasks as messages to the edge level and implement storage and computation processes. Resources, which include CPU resources $ U $, memory $ C $ and transmission bandwidth $ N $, are consumed during transmission and computation. The latency of transmission and computation in the offload process is define as $ LA $. Note that each edge server region, based on priority, contains a message queue $ Mq $. To help understand the overall process, major notations are summarized in Table~\ref{tab1}.

The mobile application is initially divided into several subtasks using segmentation algorithms based on the characteristics of the mobile application, such as considering the functional and nonfunctional partition. Then, depending on the available computing resources of the mobile device, some tasks are executed immediately to obtain the result. However, most tasks are transmitted to nearby edge servers in a uniform manner. As mentioned above, a message queue is used for task storage: tasks are stored in the server region in the form of a queue, and each task is allocated to a corresponding edge server based on the task classification to perform offloading. Different tasks have different processing priorities: if the subtasks are uniformly offloaded to the edge servers, a priority problem will arise. Thus, subtasks are given different priorities. As part of the improved design, a priority task scheduling algorithm is implemented at the edge server to reduce the delay and the prioritize requests. The following sections discuss how to model the system,which is a precondition to use reinforcement learning.
\begin{table}
\centering  
\caption{The Notations for MEC Framework.}
\label{tab1}
\begin{tabular}{ll}
\hline
\hline
Symbol           & Definition \\
\hline
$ D $            & Cloud data center \\
$ E $            & Edge server set \\
$ M $            & Mobile device set \\
$ m $            & Number of edge servers \\
$ n $            & Number of mobile devices \\
$ C_i^{in} $     & Input data size of the $ i_{th} $ subtask \\
$ N_i^{dow} $    & Downlink bandwidth of the $ i_{th} $ subtask \\
$ U_i $          & CPU resources required for the $ i_{th} $ subtask deployment \\
$ C_i^{out} $    & Output data size of the $ i_{th} $ subtask \\
$ N_i^{up} $     & Uplink bandwidth of the $ i_{th} $ subtask \\
$ P_i $          & Priority of the $ i_{th} $ subtask \\
$ V_i $          & CPU utilization of the $ i_{th} $ computing device \\
$ LA^{trans} $   & Data transmission latency of the $ i_{th} $ subtask  \\
$ LA^{comp} $    & Computation latency of the $ i_{th} $ subtask   \\
$ ST $              & the service times matrix for all subtasks \\
\hline
\hline
\end{tabular}
\end{table}

\subsection{Task Dependency Model}
We call tasks divided by a segmentation algorithm subtasks. A portion of the subtasks, such as user interaction tasks and device I/O tasks, can be locally processed. Another portion, especially computation tasks with large amounts of data, can be offloaded to an edge server~\cite{lu2020optimization}. Although these tasks have data-dependency on each other, they can still be executed on different devices, which makes the fine-grained offloading decisions possible.
\begin{figure}
\centering 
\includegraphics[scale=0.5]{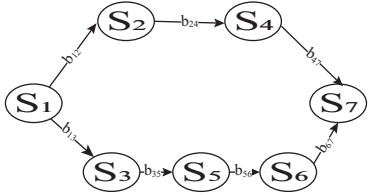}
\caption{Example of the data dependency of subtasks.} \label{dependecy}	
\end{figure}

The divided subtasks can be represented as a directed acyclic graphic (DAG) $ g=(S,B) $. Each node $ s_i \in S $ of the DAG represents one subtask, and each edge $ b_{ij} \in B $ represents the data dependency between tasks such that task $ s_j $ should receive the result of task $ s_i $ before its execution. As shown in Fig.~\ref{dependecy}, the set of subtasks after dividing a mobile application is $ S=\{s_1,s_2,s_3,s_4,s_5,s_6,s_7\} $. Tasks $ s_1 $ and $ s_7 $ should be executed on the local device, and the remaining subtasks can be offloaded as needed.

\subsection{Task Priority Model}
The analytic hierarchy process (AHP) model is used to determine the task priority. AHP, which has been applied in various fields, is method suitable for solving priority-based scheduling problems~\cite{zeleny2012multiple,zanakis1998multi,saaty1988analytic}. During the time interval $ 0 \sim t $, let $ M=\{Mq_1, Mq_2, \dots, Mq_j\}$ be the task sequence generated by each mobile device. To determine the priority of tasks, the data size of the message, the CPU cycles required by the task, and the deadline are considered. In our study, the order of importance of these factors is $<$Deadline, CPU Cycle, Data Size$>$. The deadline has the highest weight when describing the division of priorities.
 
First, the factors of the same level are compared and constructed into the analytic hierarchy matrix $ A=(a_{ij})_{3 \times 3} $:
\begin{equation}
a_{ij}=\frac{1}{a_{ji}} 
\end{equation}

where $ a_{ij} $ is the result of comparing the importance of factor $ i $ and factor $ j $. Table~\ref{tab2} shows different importance levels and their weights~\cite{dos2019analytic}.

\begin{table}[!t]
\centering  
\caption{The major importance factors.}\label{tab2}

\begin{tabular}{ll}
\hline 
\hline
factor i vs factor j & weight \\
\hline
Equally Strong & 1       \\
Weakly Stronger  & 3       \\
Stronger       & 5       \\
Much Stronger  & 7       \\
Absolution   & 9       \\
Other        & 2,4,6,7 \\
\hline 
\hline
\end{tabular}
\end{table}

Then, the matrix of weights of all tasks $ \Delta =(u_r^k)_{3 \times J} $ is constructed, where $ J $ represents the number of tasks and $ u_r^k $ represents the weight of the $ r_{th} $ task based on the $ k_{th} $ factor:
\begin{equation}
U_r^k=\frac{\sum_{j=1}^{n}a_{rj}}{\sum_{i=1}^{3}\sum_{j=1}^{3}a_{ij}} 
\end{equation}

Finally, the priority vector $ PV $ of each task is generated. $ PV= \Delta \times \Lambda $, where $ \Lambda $ is the eigenvalue of its weight according to the AHP matrix.

\subsection{Energy Consumption Model}
The limited battery capacity of mobile devices requires optimization to minimize energy consumption. We consider the energy consumption $ {En}^{comp} $ of all devices and transmission energy consumption $ {En}^{trans} $ of mobile devices. Power consumption is correlated with the CPU usage rate~\cite{garg2013framework}. The energy consumption of the $ i_{th} $ device is as follows~\cite{zhang2017energy}:
\begin{equation}
P_{i}(u)=\left\{\begin{array}{lr}
K * P_{i}^{full} * u & if\ u>0 \\
0 & otherwise 
\end{array}\right.
\end{equation}

where $ K $ represents the ratio of idle devices to fully loaded devices, $ P_i^{full} $ represents the energy consumption in the $ i_{th} $ computing device at full-load status, and $ u $ is the CPU usage rate. The load of  computing devices varies over time. Suppose that $ u(t) $ is the CPU usage rate of the device within $ \Delta t $ time. From $ t_0 $, the continuous time computing energy consumption of the device through time $ t $ is defined as follow:
\begin{equation}
{En}^{comp}(t_0)=\int_{t_0}^{t_0+t}P_i(u(t))\,dt 
\end{equation}

Next, the data transmission rate $ r_j $ of the $ j_{th} $ mobile device in a certain channel is defined as follows~\cite{wang2017computation}:
\begin{equation}
r_j=B \log_2{(1+\frac{p_jh_j}{\sigma^2+I})}
\end{equation}

where $ B $ represents the fixed transmission bandwidth of the channel; $ p_j $ represents the energy consumption of data transmitted by the $ j_{th} $ device; $ h_j $ represents the fixed channel gain during the offloading process between the mobile device and edge server; $ \sigma^2 $ represents the noise of the mobile device; and $ I $ represents the interference power between mobile devices.

Similarly, $ f(x) $ is defined as the energy consumption $ p_j $ of data transmission in the $ j_{th} $ mobile device. The transmission energy consumption of the $ j_{th} $ mobile device during time unit $ t $ is then calculated as follows:
\begin{equation}
{En}_j^{trans}(t_0)=p_jt_0=f(r_j)t_0 
\end{equation}

Suppose that the total number of devices is $ (1+m+n) $, which includes $ n $ mobile devices, $ m $ edge servers and one cloud data center. Finally, the total energy consumption during time unit $ t $   is defined as follows:
\begin{equation}
{En}_{sum}(t_0)=\sum_{i=1}^{1+m+n}{En}_i^{comp}(t_0)+\sum_{j=1}^{m}{En}_j^{trans}(t_0)
\end{equation}

\subsection{Average latency}
Here, we consider the dependency model between subtasks. The average latency ${LA}^{avg} $ of all mobile applications consists of two parts: 1) the data transmission time ${LA}^{trans} $ between subtasks; 2) the data computation time ${LA}^{comp} $ for subtasks. Since the data transmission volume and available bandwidth may change over time, the general transmission latency of the $ i_{th} $ subtask for downloading dependent data and uploading results during time unit $ t $  is defined as follows:
\begin{equation}
\left\{\begin{array}{lr}
LA_i^{dow}(t_0)= \int_{t_0}^{t_0+t}\frac{c_i^{in}(t)}{N_i^{dow}(t)}\,dt \\
\,\\
LA_i^{up}(t_0)= \int_{t_0}^{t_0+t}\frac{c_i^{out}(t)}{N_i^{up}(t)}\,dt
\end{array}\right.
\end{equation}

Then, the data transmission latency for each subtask during time unit $ t $   is defined as follows:
\begin{equation}
{LA}_{i}^{trans}(t_0)= x_{dow} {LA}_i^{dow}(t_0)+ x_{up}{LA}_i^{up}(t_0)
\end{equation}

where $ x_{dow} $ indicates whether the current subtask needs to download dependent data through the network and $ x_{up} $ indicates whether the current subtask needs to upload the processing result to the network. 

In addition, the computation latency of the $ i_{th} $ subtask consumed by processing data during time unit $ t $  is defined as follows:
\begin{equation}
\begin{split}
{LA}_{i}^{comp}(t_0)=x_{off}\int_{t_0}^{t_0+t}{\frac{c_i^{in}(t)}{f^{server}(t)}} \, dt \\ +
(1-x_{off})\int_{t_0}^{t_0+t}{\frac{c_i^{in}(t)}{f^{local}(t)}} \, dt
\end{split}
\end{equation}

where $ f^{server}(t) $ represents the processing rate of the edge server for deploying the $ i_{th} $ subtask; $ f^{local}(t) $ represents the processing rate of the local device for deploying the $ i_{th} $ subtask; and $ x_{off} $ indicates whether the current subtask is offloaded to the edge server for execution. 

The average latency of all mobile applications in time unit $ t $ is defined as follows:
\begin{equation}
{LA}^{avg}(t_0)=\frac{ \sum_{j=1}^{AN} \sum_{i=1}^{{TN}_j}({LA}_{i}^{trans}(t_0)+{LA}_{i}^{comp})}{AN}
\end{equation}

where $ AN $ represents the number of applications and $ {TN}_j $ represents the number of subtasks after the $ j_{th} $ application is divided.

\subsection{Load status}
By calculating the load status of each computing device to represent the utilization status of the cluster,  server overload can be effectively avoided and the network data processing ability can be enhanced. Eq.(12) is used to calculate the load of all resources in the $ i_{th} $ computing device $ Load_i(t_0) $ during $ t_0 $~\cite{randles2010comparative}:
\begin{equation}
Load_i(t_0)=\int_{t_0}^{t_0+t}(\sum_{k=1}^{ls} U_k \times L_k(t) \, dt )
\end{equation}

where $ ls $ represents the number of indicators used to calculate the load status; $ U_k $ is the weight of each resource; $ L_k(t) $ represents the usage rate of each resource per unit time. Then, the comprehensive load status $ LS(t_0) $ of all computing nodes is calculated using the root mean square error (RMSE) method. The smaller the value is, the better the load status result.

\subsection{Network usage}
Network usage is the amount of data transmitted by all mobile devices in time unit $ t $. Excessive network usage will cause network congestion, which will reduce the performance of the entire offloading process. The formula of network usage per unit time is as follows:~\cite{ferris2017adjusting}:
\begin{equation}
\left\{\begin{array}{lr}
TD(t_0) = \sum_{j=1}^{AN} \sum_{i=1}^{TN_j} \int_{t_0}^{t_0+t}(x_{dow}C_i^{do}(t_0)
+x_{up}C_i^{up}(t_0)) \\
\,\\
\quad \quad \quad  NU(t_0) = \frac{SL^{avg}(t_0) \times AN \times TD(t_0)}{t}
\end{array}\right.
\end{equation}

where $ TD(t_0) $ represents the total amount of data transmission generated by all mobile applications during time $ t_0 $ and $ NU(t_0) $ represents the network usage during time $ t_0 $.

\section{The Algorithm and Our Optimization}
\begin{figure}
\centering 
\includegraphics[scale=0.4]{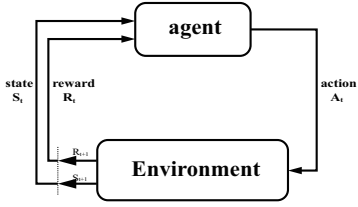}
\caption{Reinforcement learning paradigm.} \label{dql}
\end{figure}

The general reinforcement learning model is shown in Fig.~\ref{dql}. In each time step $ t $ during the learning process, the agent observes state $ s_t $ and takes action $ a_t $ based on current policy $ \pi $. When the state of the environment changes to $ s_{t+1} $,  a reward value $ r_t $ is received in the next time step. The state transition of the environment and the rewards obtained have the Markov feature, that is, the probability and rewards of state transition depend only on the state of the environment $ s_t $ and the action $ a_t $~\cite{li2019deepjs}. The agent receives these quantities according to the decision policy, interacting with the environment and backpropagation, to maximize the expectation of reward $ \mathbb{E}[\sum_{t=1}^T\,r_t] $~\cite{sutton2018reinforcement}.

\subsection{State Space}
The reinforcement learning model is constructed by combining the features between subtasks and servers in MEC. According to the above section and notation in Table 1, the state space can be described as a tuple $ s_t=(\{C_i^{in}\}_{i=1}^{T},\{N_i^{dow}\}_{i=1}^{T}, \{U_i\}_{i=1}^{T},P,\{C_i^{out}\}_{i=1}^{T},\{N_i^{up}\}_{i=1}^{T}, L,\\ \{V_i\}_{i=1}^{n+m}) $, where $ C_i^{in} $ represents the amount of input data required for the $ i_{th} $ subtask; $ N_i^{dow} $ represents the downlink bandwidth that the $ i_{th} $ subtask needs; $ U_i $ represents the CPU resources required for deployment of the $ i_{th} $ subtask; $ P $ is the priority matrix for all subtasks; $ C_i^{out} $ represents the amount of results data generated by the $ i_{th} $ subtask; $ N_i^{up} $ represents the uplink bandwidth that the $ i_{th} $ subtask needs for uploading the data; and $ V_i $ represents the CPU usage rate of the $ i_{th} $ device at time step t. In the mobile environment, the subtask offloading decision involves a total number of $ (n+m) $ computing devices, including $ n $ mobile device and $ m $ edge servers.

\subsection{Action Space}
To ensure that the agent is able to choose an appropriate computing device, one-to-one correspondence with the set of computing devices and tasks is mapped as the action space~\cite{deng2020dynamical}. It indicates that the computing device satisfies the task requirements. The size of the action space is $ (n+m+1)^K $, where the $ 1 $ means that the agent decides to locally execute the task. However, the large action space in early iterations makes it difficult for the agent to learn effective decisions, making it difficult for the model to converge. Therefore, preprocessing of the action space helps to effectively learn actions in the iterative process, thereby reducing the number of iterations. $ A_{valid}[i] $ indicates whether the $ i_{th} $ device can be used as the target device for task offloading. $ A_{valid}[i] $ is 1 if the resources required by the task can be satisfied by the available resources of the target computing device; otherwise, it is 0. The valid action space algorithm is defined in Algorithm~\ref{a1}.

\begin{algorithm}
	\caption{Valid action space algorithm.}
	\label{a1}
	\LinesNumbered
	\KwIn{The subtask $ T $ that needs to be offloaded, all edge servers $ serverList $, including including the local mobile device and edge servers.}
	\KwOut{The valid action space $ A_{valid} $.}
	\textbf{Initialize} the valid action space $ A_{valid} $ and $ size $ = size of $ A_{valid} $\; 
	\textbf{Initialize} the $ Re_{re} $, which is the resources needed for  task $ T $ deployment\;
	\For{i=1:size-1}{	
		$ Re_{av} $=the available resources of current device $ serverList[i] $\;
		\eIf{$ Re_{re} > Re_{av} $}{\
			$ A_{valid}[i]=0 $;
		}{
			$ A_{valid}[i]=1 $;}
	}
\end{algorithm}

\subsection{Reward Function}
The reward function  guides the learning process of the agent. Different performance optimization targets require different reward functions. In our study, the reward of the offloading policy is assessed according to three factors: energy consumption, average latency and load status of all devices in MEC. The reward function for time $ t_0 $ is defined as follows:
\begin{equation}
\left\{\begin{array}{lr}
R = -(\alpha Z_{en}(t_0)+\beta (Z_{la}(t_0)+P)+\gamma Z_{ls}(t_0)) \\
\quad \quad \quad \quad \quad \quad \quad \alpha+\beta+\gamma=1
\end{array}\right.
\end{equation}

where $ Z_{en}(t_0) $, $ Z_{la}(t_0) $, $ P $ and $ Z_{ls}(t_0) $ represent the energy consumption, average latency, task priority and load status of all devices at $ t_0 $ calculated by Eq.(7), Eq.(11), Eq.(2) and Eq.(12) in Section 3. These values are normalized using z-score standardization.

The action-value function used in reinforcement learning algorithm describes the expected return on following policy $ \pi $ in time step $ t $~\cite{lowe2017multi}:
\begin{equation}
\begin{split}
Q^\pi (s_t,a_t)=\mathbb{E}_{r_t,s_{t+1} \sim E }[r_t(s_t,a_t)\\+\gamma \mathbb{E}_{a_{t+1} \sim \pi} [Q^\pi (s_{t+1},a_{t+1})]]
\end{split}
\end{equation}

Thus, the target policy is described as a function $ \mu :S \leftarrow A $ if the update of the target policy is continuous:
\begin{equation}
\begin{split}
Q^\mu (s_t,a_t)=\mathbb{E}_{r_t,s_{t+1} \sim E }[r_t(s_t,a_t)\\+\gamma \mathbb{E}_{a_{t+1} \sim \pi} [Q^\pi (s_{t+1},\mu (s_{t+1}))]]
\end{split}
\end{equation}

\subsection{Algorithm Optimization}

\subsubsection{Optimization based on LSTM}
The MEC environment has limited cognitive ability, which means it is difficult to directly observe the underlying state of the system in the current time step. A common approach is to use the partial observation Markov decision process (POMDP) to model a system and help make decisions with incomplete state information~\cite{aviv2005partially}. 

The POMDP is defined as a 7-tuple $ (S,A,T,R,\Omega,O,\gamma) $, where $ S $ represents the state set of the environment; $ A $ represents a set of actions; $ \Omega $ represents a set of observations; $ T:S \times A \rightarrow \pi(S) $ is a set of conditional transition probabilities between states; $ R:S \times A \rightarrow R $ is a reward function; $ O:S \times A \rightarrow \pi(Z) $ is a set of conditional probabilities; and $ \gamma \in [0,1] $ is the discount factor. Although DDPG achieves good results when the agent obtains complete observations from a real environment, the fact that the state information of a system cannot be directly observed and is only partially known complicates MEC.

\begin{figure}
	\centering
	\includegraphics[scale=0.4]{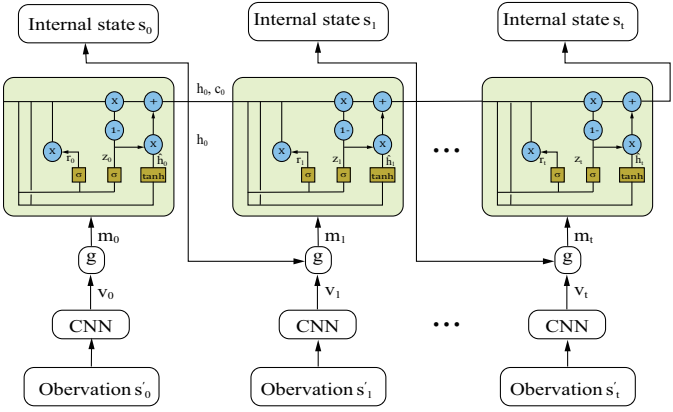}
	\caption{LSTM-based internal state prediction.} \label{LSTM}
\end{figure}

To address the dynamic changing feature, LSTM and DDPG are integrated to enhance the task offloading method. We add an LSTM layer to the DDPG network to accurately estimate the current state. As shown in Fig.~\ref{LSTM}, the internal state prediction layer consists of three types of networks: convolutional neural network (CNN), attention network and recurrent network (LSTM). At each step time $ t_0 $, the CNN receives the current state space $ s_t^{\prime} $, also called observation, to extract the environment feature. The attention network inputs a set of vectors $ v_t $ from the CNN and outputs a context vector $ m_t $ as a combination of the input. The LSTM takes the context vector and the previous hidden state $ h_{t-1} $ and memory state $ c_{t-1} $ and produces hidden state $ h_t $ and memory state $ c_t $. The hidden state $ h_t $ is then used to evaluate the underlying state $ s_t $.

\subsubsection{Optimization from multiagent communication}
The deterministic policy gradient consists of two kinds of networks: policy network (actor) and Q-network (critic). The former is responsible for selecting the current action according to the current state; the latter is responsible for calculating the current Q value $ Q^\mu (S, A) $. Although Eq.(16) is applicable to single-agent methods, such as deterministic policy gradient, these approaches lack the principle mechanism to promote team collaboration~\cite{qiu2019deep}. A multiconnection network based on the BRNN enables agents to communicate with each other before taking actions, which is the starting point of our innovation.

\begin{figure}
	\centering
	\includegraphics[scale=0.5]{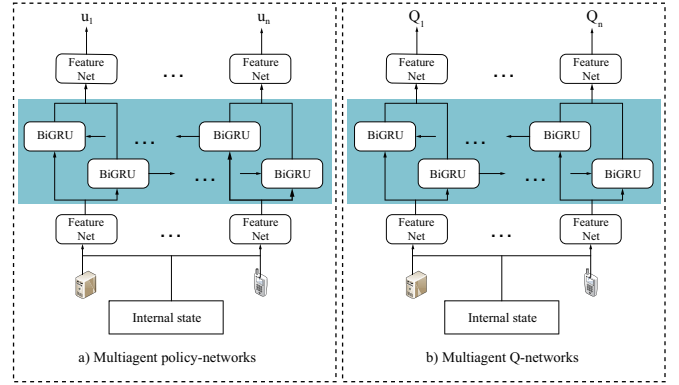}
	\caption{Mutual communication network.} \label{comm}
\end{figure}

As shown in Fig.~\ref{comm}, the proposed approach combines a deep deterministic policy gradient with the multiconnection network to enable agents to communicate with each other before taking actions. The network receives the internal state from the internal state prediction unit. In contrast to the previous deep deterministic policy gradient algorithm, the network shares the observation state of each agent before the output by the multiconnection network. The multiconnection network consists of a multiagent policy network and multiagent Q-network: both the policy network and the Q-network are based on BRNNs. As a means of communication, BRNNs have been used to connect each individual agent's policy and Q-networks. The policy network takes observations of other agents and local observations as inputs and determines all actions for the agents.

The backward gradient of all networks applies backpropagation through time (BPTT)~\cite{werbos1990backpropagation}. In other words, gradients derived from all agent rewards are first corrected by propagation. Then, returned gradients can be further propagated back to update the related parameters. Combined with Eq.(16),  $ J_i(\theta)=\mathbb{E}_{r_t,s_{t+1} \sim E }[r_t(s_t,a_t)+Q^{\pi_\theta}(s_{t+1},\mu(s_{t+1}))] $ is used to describe the rewards of the $ i_{th} $ agent. Thus, the reward of all agents, denoted by $ J(\theta), $ is defined as follows:
\begin{equation}
J_i(\theta)=\mathbb{E}_{r_t,s_{t+1} \sim E }[\sum_{i=1}^N r_i(s,\pi_\theta(s))]
\end{equation}

To perform gradient descent, the gradient of $ J(\theta) $ is calculated by the following rule:
\begin{equation}
\nabla J(\theta)=\mathbb{E}_{r_t,s_{t+1} \sim E }[\sum_{i=1}^N \sum_{j=1}^N \nabla_\theta \pi_{j,\theta}(s) \times \nabla_{\pi_j} Q_i^{\pi_\theta}(s,\pi_\theta(s)) )]
\end{equation}

Here, the parameters of networks are shared among agents and uses stochastic gradient descent (SGD) to optimize the policy-network and Q-network. Since the number of parameters is independent of the number of agents to accelerate the learning process and enable domain adaption, where the extended training of networks is performed from a small group to a larger group of agents~\cite{peng2017multiagent}. In addition, parameter sharing is suitable for the cooperation of MEC scenarios.

\subsubsection{Com-DDPG algorithm}
The above optimization methods make our Com-DDPG algorithm suitable for task offloading in MEC scenarios. To train the algorithm, we first store transitions in the replay buffer, including the current state, current action, reward and next state of each agent. Then, we sample a random mini-batch of transitions from the replay buffer. Finally, we compute the gradient estimation of the critic and actor to update the networks based on SGD. The pseudo code of the Com-DDPG algorithm in each sampling process is shown in Algorithm~\ref{a2}.

\begin{algorithm*}
	\caption{Com-DDPG algorithm}
	\label{a2}
	\LinesNumbered
	\textbf{Initialize} actor network and critic network with $ \xi $ and $ \theta $\; 
	\textbf{Initialize} target network and critic network with $ \xi^\prime \leftarrow \xi $ and $ \theta^\prime \leftarrow \theta $\;
	\textbf{Initialize} replay buffer $ R $\;
	\For{episodes=1:$ E $}{	
		\textbf{Initialize} a random process $ \upsilon $ for action exploration\;
		Receive initial observation $ S $\;
		\For{t=1:$ T $}{
			\For{ each agent $ i $, select and execute action $ a_i^t $}{
				receive reward $ \{r_i^t \}_{i=1}^N $ and observe new state $ s^{t+1} $ \;
				store transition $ \{s^t, \{a_i^t \}_{i=1}^N, \{r_i^t \}_{i=1}^N, s^{t+1} \} $ in $ R $ \;
				sample a random mini-batch of M transitions:\
				$ \{ s_m^t, \{a_{m,i}^t \}_{i=1}^N, \{r_{m,i}^t \}_{i=1}^N, \{s_m^{t+1}\}_{m=1}^M \} $ from $ R $ \;
				compute target value $ \hat{Q}_{m,i} $ for each agent in $ R $\;
				compute critic gradient estimation $ \Delta \xi $ \;
				compute actor gradient estimation $ \Delta \theta $ \;
				update the networks based on SGD using the above gradient estimators\;
				update the target networks: \newline
				$ \xi^\prime \leftarrow \gamma \xi + (1-\gamma) \xi^\prime, 
				\theta^\prime \leftarrow \gamma \theta + (1-\gamma) \theta^\prime $\;
			}
		}
	}
\end{algorithm*}

\section{Experiment}
In this section, we simulate the task offloading problem based on MEC. The performance of the offloading decision is assessed by comparing the energy consumption, load status, execution latency and network usage generated by each algorithm in the MEC cluster. The following simulation algorithms are considered: 1) the local strategy based on preference for local devices; 2) the edge first strategy based on a first-fit algorithm; 3) the DQN strategy based on deep reinforcement learning; 4) the improved deep reinforcement learning strategy (DRQN) based on LSTM; and 5) the Com-DDPG strategy based on multiagent cooperation and LSTM.

\subsection{Data Preprocessing and Parameter Setting}
The cluster comprises a cloud data center, 80 edge servers and multiple mobile devices. The edge servers are divided into 10 regions by relative distance to the server node, and only one offloading request per mobile device is made within the unit time. In our study, we use log data from Alibaba Cluster Data V2018 \footnote{https://github.com/alibaba/clusterdata} to simulate the task dependence the offloading process. A total of 30,756 tasks (5,000 jobs, 2,515,063 task instances) are selected from the cluster data to train the networks; then, 100 jobs are randomly selected from the remaining data to test the efficiency of the strategy. The jobs and tasks represent tasks and subtasks in the MEC environment. Moreover, the parameters of all reinforcement learning algorithms are the same to ensure credible training results. The learning rate of SGD is $ \alpha = 0.005 $, the batch size  is $ K = 16 $, the epoch period is $ C = 50 $, and the discount rate is $ \gamma = 0.9 $. For the LSTM based DRQN algorithm, the time window $ W $ is set to 10. The relevant parameters are presented in Table~\ref{tab3}.

\begin{table}
	\caption{Parameters used in the experiments.}
	\label{tab3}
	\renewcommand\arraystretch{1.5}
	\begin{tabular}{m{1.5cm}<{\centering}m{1.5cm}<{\centering}m{3.7cm}<{\centering}}
	\hline
	\hline
	Parameter    & Fixed value & Description \\
	\hline
	jobs number & 5,000       & the number of jobs for training \\
	$ \gamma $   & 0.005       & the learning rate of SGD \\
	$ \alpha $   & 0.9         & the discount rate \\
	$ K $        & 16          & the batch size of learning \\
	$ C $        & 50          & the epoch period \\
	$ W $        & 10          & the time window of LSTM  \\
	$ B $        & 1MHZ        & the fixed transmission bandwidth of the channel \\
	$ h_j $      & -50db       & the channel gain of the $ j_{th} $ mobile device \\
	$ \sigma^2 $ & -100dBm     & the channel noise power \\
	\hline
	\end{tabular}
\end{table}

\subsection{The loss function experiment}
In machine learning, loss functions represent the price paid for inaccurate learning results. The smaller the loss function value is, the better the result of the network model. The loss function is defined as the absolute value of the difference between the reward generated by the algorithm and the reward calculated from the data sets. The algorithms considered for comparison are DQN, DRQN and Com-DDPG, which are the reinforcement learning algorithms.

\begin{figure}
\centering
\includegraphics[scale=0.4]{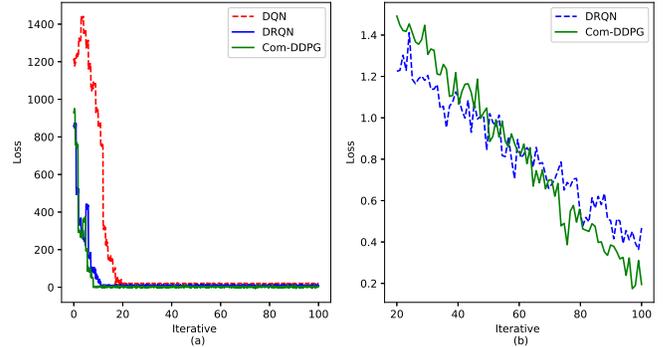}
\caption{Iterative figure of loss functions for Com-DDPG, DRQN and DQN.} \label{loss}
\end{figure}

 Fig.~\ref{loss} presents the loss function scores for the first 100 iterations of the deep reinforcement learning algorithm. In Fig.~\ref{loss} (a), the loss function scores for the three algorithms show a decreasing trend. After approximately 20 iteration, all three algorithms tend to converge. However, the DRQN and Com-DDPG algorithms have lower loss function scores than the DQN algorithm at a given iteration. Since the loss function scores of DQRN and Com-DDPG are similar,  detailed  scores are shown in Fig.~\ref{loss} (b). At the initial iteration, the loss function score of the Com-DDPG algorithm is similar to that of the DRQN algorithm. With increasing iteration, network learning shares more information from the multiconnection unit, and the loss function score of the Com-DDPG algorithm experiences a greater decrease  than that of the DRQN algorithm. Furthermore, comparing the DQN algorithm with the DRQN and Com-DDPG algorithms, the deep reinforcement learning algorithm score based on the LSTM network is relatively low during the entire iteration process. The main reason is that the DRQN algorithm and Com-DDPG algorithm obtain more information about the underlying state in the LSTM network layer. Therefore, our Com-DDPG algorithm is better able to approach the optimal solution as the number of iterations increases.

\subsection{The maximum completion time experiment}
The maximum completion time is defined as the time of task completion minus the time of task submission. All jobs are divided into several blocks according to 10 consecutive jobs; then, each block is trained from left to right. Because each job contains a different number of tasks, the number of tasks changes over time.

\begin{figure} 
\centering
\includegraphics[scale=0.5]{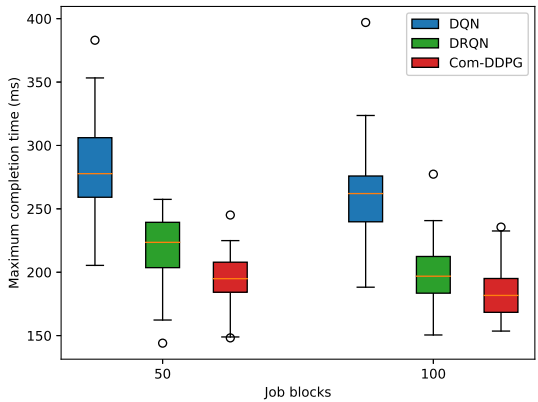}
\caption{Maximum completion time for different algorithms.} \label{max}
\end{figure}

The maximum completion time between different reinforcement learning algorithms is shown as box plot in Fig.~\ref{max}. The maximum completion time is generally decreasing since each algorithm adjusts its parameters for task offloading during the training process. On average, the DRQN algorithm reduces the maximum completion time by approximately 17\%, and our Com-DDPG algorithm reduces the time by approximately 23\%. The maximum completion time of the DQN algorithm is within the range of approximately $ 250 ms\sim310 ms $, that of the DRQN algorithm is within the range of $ 195 ms\sim220 ms $, and that of the Com-DDPG algorithm is within the range of $ 160 ms\sim200 ms $. Therefore, the scheduling scheme given by Com-DDPG and DRQN has a more compact distribution because the Com-DDPG and DRQN algorithms obtain more environment information from the LSTM module to make offloading decisions. In addition, the Com-DDPG algorithm has fewer outliers because of the closer cooperation among the  agents during the training process. Therefore, Com-DDPG has better stability and robustness than the other algorithms.

\subsection{The service times experiment}
Considering the service times, we simulate 100 jobs to carry out offloading, and monitor the run time each server consumed by each algorithm, including Edge, DQN, DRQN and Com-DDPG. The symbol $ st_i $ is defined as $ i_{th} $ server’s service times to execute subtasks.

\begin{figure*} 
\centering
\includegraphics[scale=0.5]{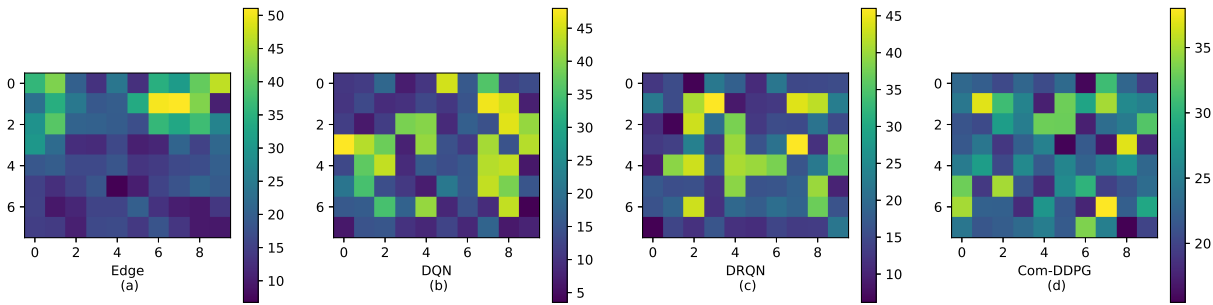}
\caption{Service times for diferent algorithms.} \label{usage}
\end{figure*}

As shown in Fig.~\ref{usage}, the results to each algorithm are shown as heat maps. In Fig.~\ref{usage}(a), the offloading target is approximately concentrated on the first 30 servers, values $ \sim >40 $. That is because Edge algorithm use first fit algorithm to choice target server. The Fig.~\ref{usage}(b) and (c) are similar. They both have some server been frequently choice, values $ \sim 40 $. The reason is that those reinforcement learning algorithm trend select server they used before. From Fig.~\ref{usage}(d) generated by Com-DDPG, although some servers are frequently selected, the number of those servers is lower than the other reinforcement learning. The main reason is that Com-DDPG based on multi-agent communication supports offloading decision before sharing their information, and makes our algorithm better to use the feature of MEC environment.

\subsection{The different numbers of jobs experiment}
In real-world scenarios, MEC environments often need to process continuous offloading requests from users. Continuously submitting jobs to the MEC will directly reflect the performance of the offloading strategy. According to Eq.(7), Eq.(11), Eq.(12) and Eq.(13), the performance of each strategy is evaluated in terms of energy consumption, load status, latency and network usage.
\begin{figure*}[htbp]
\centering
\includegraphics[scale=0.5]{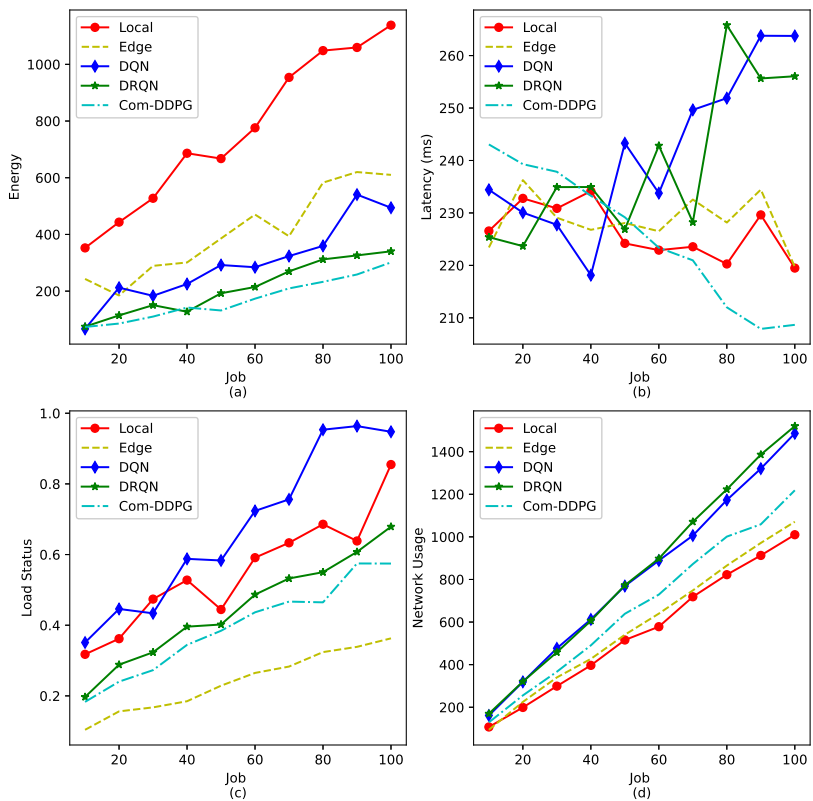}
\caption{Resource consumption generated by each algorithm in task offloading.} \label{plot4}
\end{figure*}

Fig.~\ref{plot4} is a diagram of the energy consumption, load status, latency and network usage generated by continuously executing the offloading decision of 100 tasks. As the number of jobs increases, the energy consumption, load status, latency and network usage increase. The local algorithm achieves better performance in terms of latency and network usage, but the energy consumption is the worst. This is mainly because the local algorithm executes tasks on local devices, so the local algorithm has low latency and network usage. Additionally, the offloading strategy based on the edge server priority algorithm performs well in terms of load status,  network usage and most other measurement criteria. The main reason is that the edge algorithm tends to offload subtasks to edge server clusters, which improves the load status, but the performance of other aspects is worse than that other algorithms. Meanwhile, edge server performance meets the requirements of more subtasks, thus it can reduce network transmission between subtasks, as well as the network utilization of the entire cluster.

As mentioned above, the DQN, DRQN  and Com-DDPG algorithms use deep reinforcement learning to automatically learn the offloading strategy from self-play. The results in Fig.~\ref{plot4} show that the offloading strategy generated by DQN performs poorly in terms of load status but has stable performance in terms of the other metrics as the number of jobs increases. The DRQN algorithm has a good impact on energy consumption,  latency and network usage. These algorithms have various advantages and disadvantages. The LSTM network and multiagent collaboration algorithm make the Com-DDPG have similar performance with the DRQN algorithm in terms of energy consumption, load status and latency. When the number of jobs is large, the latency of the Com-DDPG algorithm shows a downward trend. Thus, its performance is superior to that of other deep reinforcement learning algorithms. In summary, the Com-DDPG algorithm has the best energy consumption and load status performance, but the latency is moderate. Furthermore, in contrast to other offloading schemes, the latency shows a downward trend as the number of jobs increases.

\section{Conclusions and future work}
In this paper, the Com-DDPG method is proposed to implement offloading for MEC, where computation-intensive and time-sensitive applications call for rapid data processing. We study the problems of server clusters and multidependence for mobile computing tasks. Multiagent reinforcement learning considers the energy consumption, load status, execution latency and network usage as inputs and then outputs the offloading strategy. As optimization steps, BRNN and LSTM are used to learn communication features via neighbor agents and to observe the internal state to support decision making.  

In the future, we will further optimize the performance of the proposed method, for example, by adding more behavior features of mobile devices and analyzing the historical log data of the edge servers. Moreover,  we will use formal methods to verify the returned offloading strategy generated by our Com-DDPG method from the quantitative and qualitative perspectives.

%

\bibliographystyle{IEEEtran}
\bibliography{ref}


%

\begin{IEEEbiography}[{\includegraphics[width=1in,height=1.25in,clip,keepaspectratio]{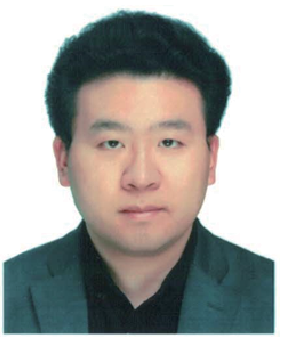}}]{Honghao Gao}
(Senior Member, IEEE) received the Ph.D. degree in Computer Science and started his academic career at Shanghai University in 2012. Prof. Gao is currently with the School of Computer Engineering and Science, Shanghai University, China. He is also a Professor at Gachon University, South Korea. Prior to that, he was a Research Fellow with the Software Engineering Information Technology Institute of Central Michigan University (CMU), USA, and was also an Adjunct Professor at Hangzhou Dianzi University, China. His research interests include Software Formal Verification, Industrial IoT/Wireless Networks, Service Collaborative Computing, and Intelligent Medical Image Processing. He has publications in IEEE TII, IEEE T-ITS, IEEE IoT-J, IEEE TNSE, IEEE TCCN, IEEE/ACM TCBB, ACM TOIT, ACM TOMM, IEEE TCSS, IEEE TETCI, IEEE JBHI, IEEE Network, and IEEE Sensors Journal.

Prof. Gao is a Fellow of IET, BCS, and EAI, and a Senior Member of IEEE, CCF, and CAAI. He is the Editor-in-Chief for International Journal of Intelligent Internet of Things Computing(IJIITC), Editor for Wireless Network(WINE) and IET Wireless Sensor Systems(IET WSS), and Associate Editor for IET Software, International Journal of Communication Systems(IJCS),  Journal of Internet Technology(JIT), and Journal of Medical Imaging and Health Informatics(JMIHI). Moreover, he has broad working experiences in industry-university-research cooperation. He is a European Union Institutions appoint external expert for reviewing and monitoring EU Project, is a member of the EPSRC Peer Review Associate College for UK Research and Innovation in the UK, and is also a founding member of IEEE Computer Society Smart Manufacturing Standards Committee.
\end{IEEEbiography}
\begin{IEEEbiography}[{\includegraphics[width=1in,height=1.25in,clip,keepaspectratio]{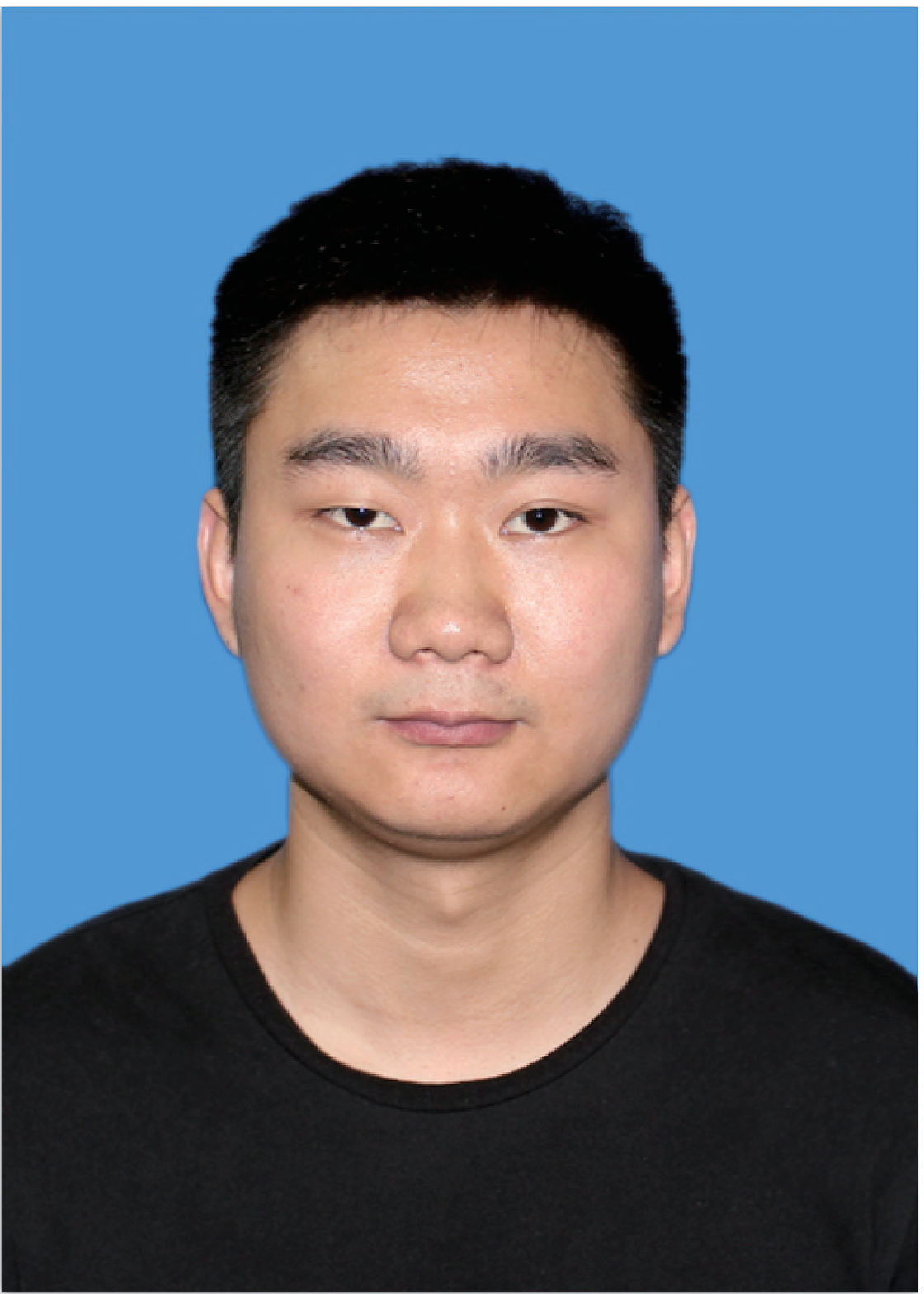}}]{Xuejie Wang}
 is currently pursuing the M.S. degree in computer science with the School of Computer Engineering and Science, Shanghai University, Shanghai, China. His research interests include edge cloud compution and reinforcement learning.
\end{IEEEbiography}

\begin{IEEEbiography}[{\includegraphics[width=1in,height=1.25in,clip,keepaspectratio]{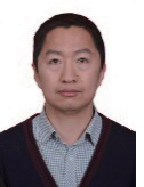}}]{Xiaojin Ma}
Xiaojin Ma received the BS, MS in Computer Science and Management Science and Engineering from Henan University of Science and Technology in 2003, 2013, respectively. He is working toward the Ph.D. degree in Shanghai University, China. His research interests include cloud computing and  parallel computing.
\end{IEEEbiography}


\begin{IEEEbiography}[{\includegraphics[width=1in,height=1.25in,clip,keepaspectratio]{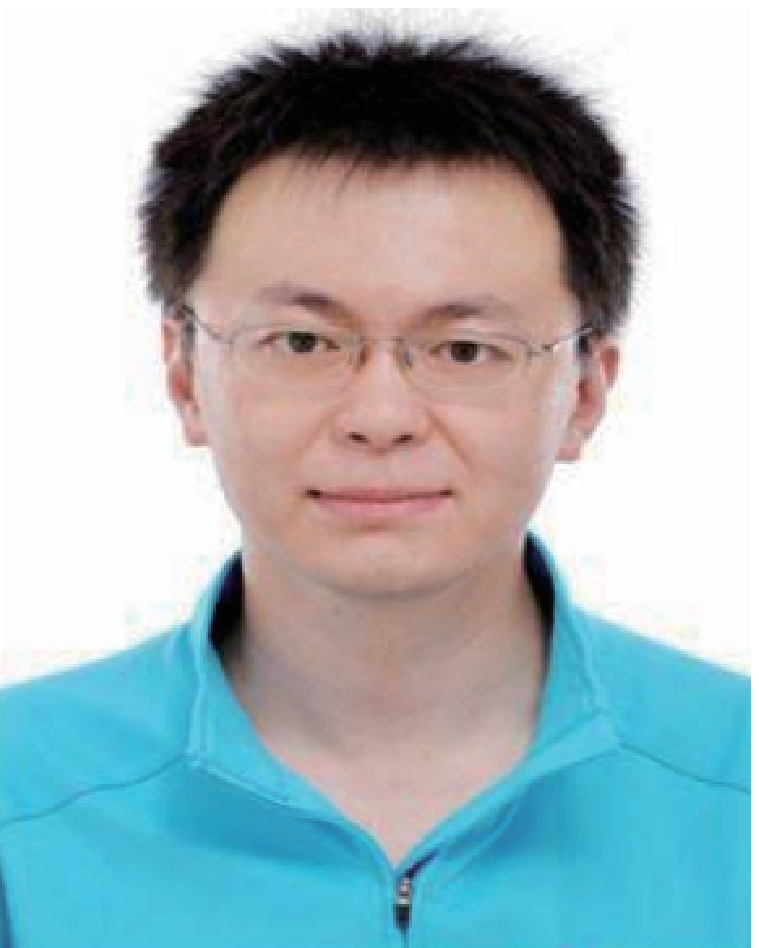}}]{Wei Wei}
 (Senior Member, IEEE) received the M.S. and Ph.D. degrees from Xi’an Jiaotong University in 2011 and 2005, respectively. He is currently an Associate Professor with the School of Computer Science and Engineering, Xi’an University of Technology, Xi’an, China. His research interest is in the area of wireless networks, wireless sensor networks applications, image processing, mobile computing, distributed computing, and pervasive computing, the Internet of Things, and sensor data clouds. He is a Senior Member of CCF.
\end{IEEEbiography}


\begin{IEEEbiography}[{\includegraphics[width=1in,height=1.25in,clip,keepaspectratio]{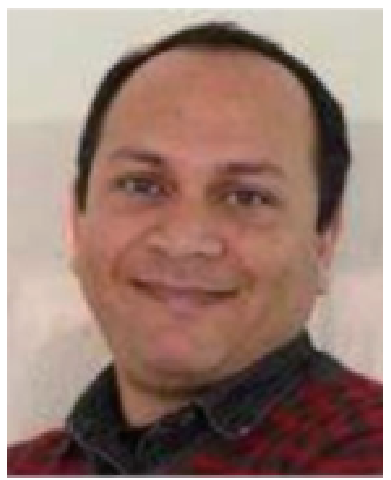}}]{Shahid Mumtaz}
(Senior Member, IEEE) received the master’s and Ph.D. degrees in electrical and electronic engineering from the Blekinge Institute of Technology, Karlskrona, Sweden, and University of Aveiro, Aveiro, Portugal, in 2006 and 2011, respectively. He has more than 12 years of wireless industry/academic experience. Since 2011, he has been with the Instituto de Telecomunicações, Aveiro, Portugal, where he currently holds the position of Auxiliary Researcher and adjunct positions with several universities across the Europe-Asian Region. He is currently also a Visiting Researcher with Nokia Bell Labs, Murray Hill, NJ, USA. He is the author of 4 technical books, 12 book chapters, and more than 150 technical papers in the area of mobile communications. Dr. Mumtaz is an ACM Distinguished Speaker, Editor-in-Chief for IET Journal of Quantum Communication, Vice Chair of Europe/Africa Region IEEE ComSoc: Green Communications and Computing society, and Vice Chair for IEEE standard on P1932.1, Standard for Licensed/Unlicensed Spectrum Interoperability in Wireless Mobile Networks.
\end{IEEEbiography}

\end{document}